\documentclass[physrev,reprint,superscriptaddress,bibnotes]{revtex4-2}  

\usepackage[T1]{fontenc}
\usepackage[australian]{babel}
\usepackage[a4paper,centering,hmargin=1.75cm,vmargin=2cm]{geometry} 

\usepackage{amsmath,amssymb,graphicx,bm,microtype}
\usepackage[dvipsnames]{xcolor} 
\usepackage{booktabs} 
\usepackage[colorlinks,allcolors=blue!50!black]{hyperref} 
\usepackage[all]{hypcap} 
\usepackage{cleveref} 
\usepackage{orcidlink}
\newcommand{\orcid}[1]{\,\orcidlink{#1}}

\usepackage{braket} 
\usepackage{siunitx,textcomp} 
\DeclareSIUnit\bohr{\text {\ensuremath {a}}_{0}}
\DeclareSIUnit\hartree{\text {\ensuremath {E}}_{\mathrm {h}}}
\DeclareSIUnit{\angstrom}{\textup{\AA}}

\newcommand{\abs}[1]{\lvert #1 \rvert}

\renewcommand{\vec}[1]{\bm{\mathrm{#1}}}

\begin{document}

\title{Delocalisation explains efficient transport and charge generation in neat Y6 organic photovoltaics}

\author{Daniel Balzer\orcid{0000-0002-4640-1059}}
\affiliation{School of Chemistry and University of Sydney Nano Institute, University of Sydney, NSW 2006, Australia}

\author{Paul A. Hume\orcid{0000-0002-7582-7155}}
\affiliation{School of Chemical and Physical Sciences, Victoria University of Wellington, Wellington, New Zealand}
\affiliation{MacDiarmid Institute for Advanced Materials and Nanotechnology, Wellington, New Zealand}
\affiliation{The Dodd-Walls Centre for Photonic and Quantum Technologies, Dunedin 9016, New Zealand}

\author{Geoffrey R. Weal\orcid{0000-0002-3477-160X}}
\affiliation{School of Chemical and Physical Sciences, Victoria University of Wellington, Wellington, New Zealand}
\affiliation{MacDiarmid Institute for Advanced Materials and Nanotechnology, Wellington, New Zealand}
\affiliation{Institute for Integrated Cell-Material Sciences (iCeMS), Kyoto University, Kyoto, Japan}

\author{Justin M. Hodgkiss\orcid{0000-0002-9629-8213}}
\affiliation{School of Chemical and Physical Sciences, Victoria University of Wellington, Wellington, New Zealand}
\affiliation{MacDiarmid Institute for Advanced Materials and Nanotechnology, Wellington, New Zealand}

\author{Ivan Kassal\orcid{0000-0002-8376-0819}}
\email[Email: ]{ivan.kassal@sydney.edu.au}
\affiliation{School of Chemistry and University of Sydney Nano Institute, University of Sydney, NSW 2006, Australia}

\begin{abstract}
Non-fullerene acceptors (NFA), such as Y6, have significantly improved the efficiency of organic photovoltaic devices (OPVs). However, the fundamental processes behind the high efficiencies of NFA devices have remained incompletely understood, with the high efficiencies persisting without the large energetic offsets often thought to be required for charge separation. Even more surprising has been the efficient charge generation in neat Y6 devices, where there is no energetic offset at all. Here, we simulate charge transport and separation in Y6 using delocalised kinetic Monte Carlo (dKMC) parameterised using atomistic calculations, thus taking into account the often-neglected ingredients of delocalisation, disorder, and polaron formation. Including delocalisation predicts higher carrier mobilities and exciton diffusion coefficients than is possible with classical simulations, bringing them into agreement with experimental values. Delocalisation also predicts higher charge-generation efficiencies in neat Y6, in agreement with experimental measurements. Finally, this work establishes dKMC as a realistic, predictive tool for understanding next-generation OPVs.
\end{abstract}

\maketitle

A central unresolved question in organic photovoltaics (OPVs) is how the photogenerated electron–hole pairs, bound by strong Coulomb forces in the low-dielectric materials, separate efficiently into free charges~\cite{Clarke2010,Few2015,Hou2018}. In organic semiconductors, with relative permittivities of~3--4, photoexcitation generates excitons whose binding energies far exceed the thermal energy. To enable charge separation, OPVs are commonly constructed from blends of donor and acceptor molecules that form heterojunctions. At these interfaces, offsets in energy levels drive the initial unbinding of excitons to form charge-transfer (CT) states with electrons and holes slightly displaced across the junction. However, even in this configuration, the Coulomb attraction between the charges still presents a major barrier to full separation. Nevertheless, in many OPVs, the fraction of excitons that dissociate into free carriers approaches 100\%~\cite{Park2009}.

Recently, understanding the charge-separation mechanism has become even more difficult, with a growing body of experimental work challenging the energy-offset framework of charge generation, especially in OPVs containing small-molecule non-fullerene acceptors (NFAs) such as Y6. Devices using Y6 and its derivatives have significantly higher efficiencies, with single-junction power-conversion efficiencies (PCEs) approaching 20\%~\cite{Sun2022,Chen2023,Jiang2024,Li2025}. There are many factors for their success, including simultaneously efficient electron and hole transport in Y6, along with its strong near-infrared absorption~\cite{Yuan2019,Guo2021}. However, the most important factor is that Y6-based devices can maintain efficient charge generation despite minimal energetic offsets at the heterojunction~\cite{Li_Chao_2021,Zhu2021,Zheng2022,Bin2016,Hou2018,Perdigon2020}. This unexpected behaviour in low-offset systems has prompted new theoretical explanations, including hybridised states with mixed exciton and charge-separated character~\cite{Qian2018,Eisner2019,Coropceanu2019,Qian2023} and the band bending at the interface due to quadrupole moments~\cite{Poelking2015,Perdigon2020,Aaladina2021,Karuthedath2021,Fu2023}. To further complicate matters, experiments indicate that charge separation can begin prior to reaching a heterojunction interface~\cite{Wang2020,Tu2020,Zhu2021,Zhang2020,Dimitriev2022,Li2023,Li2023_2}.

Most strikingly, free carriers have been detected in neat (or homojunction) Y6 films, which have no interfacial energetic gradient at all~\cite{Price2022,Zhang2022,Saglamkaya2023,Yan2023,Mcanally2025}, challenging existing charge-generation theories. Price et al.\ estimated that~60--90\% of excitons were converted to free charges in Y6 by fitting a kinetic model with a few free parameters to spectroscopic data~\cite{Price2022}. More recent studies place the charge-generation yield at the lower, but still unexpectedly high, 25\%~\cite{Mcanally2025}. The PCE of neat Y6 devices is much lower---a few percent---because of the high bimolecular recombination~\cite{Price2022,Yan2023,Saglamkaya2023,Mcanally2025}. Nevertheless, these devices confirm the possibility of charge separation without interfacial energetic offsets, which could enable efficient homojunction devices with increased morphological stability and reduced energy losses.

However, explaining efficient separation without interfacial energetic offsets remains an outstanding theoretical challenge~\cite{Hou2018,Liraz2022,Saglamkaya2023,Mcanally2025}. 
A number of hypotheses have been advanced to account for the high efficiency, including high dielectric constants and low exciton-binding energies~\cite{Zhu2021,Price2022}, delocalised intermediate states~\cite{Wang2020,Saglamkaya2023}, or energy cascades due to aggregation and morphology~\cite{Saglamkaya2023}. Recent theoretical work to test these hypotheses has focused on excited‑state energetics in Y6~\cite{Sharma2023,Giannini2024,Akram2025}, finding energy cascades driven by dielectric screening and inter‑domain electrostatics~\cite{Sharma2023}, a dense manifold of hybridised states with high binding energies~\cite{Giannini2024}, and low-energy CT states delocalised over six molecules with more modest binding energies~\cite{Akram2025}. However, charge separation is a kinetic process, and completely understanding it requires not only modelling the energetics of Y6, but also the excited‑state dynamics and charge‑transfer kinetics, for which detailed theoretical work is still lacking.

\begin{figure*}
    \centering
    \includegraphics[width=\textwidth]{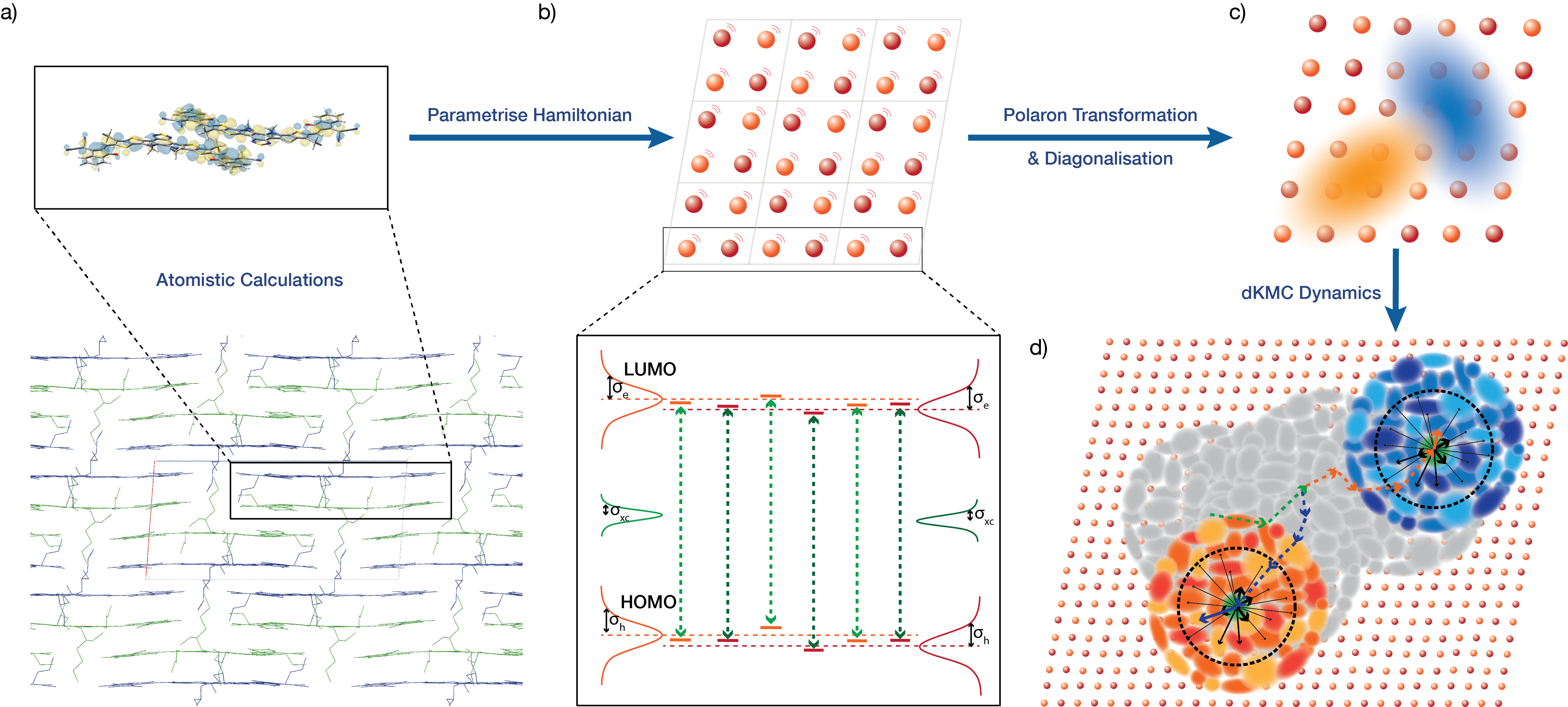}
    \caption{\textbf{Model of transport and separation processes in Y6.}
    \textbf{(a)}~Neighbouring pairs of molecules from the Y6 crystal structure are used to calculate energies of quasi-diabatic exciton and charge-transfer states, as well as couplings between them. Properties such as reorganisation energies are calculated from single-molecule geometry optimisations.
    \textbf{(b)}~The atomistic calculations are used to parameterise the dKMC Hamiltonian, in which each molecule is represented as a site located at the centroid of the molecule. The lattice is made by replicating the unit cell, each containing four sites. 
    \textbf{(c)}~The dynamics proceeds through the eigenstates of the polaron-transformed Hamiltonian. For transport simulations, these represent the location of the delocalised carrier, while for charge-generation simulations they represent the simultaneous position of the delocalised electron and hole. 
    \textbf{(d)}~Transport and separation processes are modelled using dKMC, which tracks the movement of carriers through the polaron states.}
    \label{fig:model}
\end{figure*}

Here, we show that delocalisation explains the high performance of Y6 in homojunction OPVs. To do so, we use delocalised kinetic Monte Carlo (dKMC)~\cite{Balzer2020,Balzer2022,Balzer2023,Balzer2024,Balzer2025,dKMC_github}, a simulation approach able to include the key ingredients (disorder, quantum-mechanical delocalisation, and polaron formation) in mesoscopic simulations of charge transport and separation. Our simulations are parameterised using atomistic calculations or experimental measurements of all variables that enter into the dKMC algorithm. By comparing our results with classical KMC simulations, which neglect delocalisation, we find that delocalisation improves the transport of all carriers in Y6: electrons, holes, and excitons. Delocalisation brings simulations into agreement with experimental measurements of exciton diffusion coefficients and charge-carrier mobilities. Finally, we find that delocalisation also improves charge generation efficiency in neat Y6, bringing predicted efficiencies into agreement with experiment, by producing hybridised states that mediate the transfer from excitonic to separated polaronic states.

\section{Methods\label{sec:model}}
Transport processes in Y6 are simulated using dKMC, which is detailed in previous work developing the method for charge transport~\cite{Balzer2020}, exciton transport~\cite{Balzer2023}, charge separation~\cite{Balzer2022}, and charge generation~\cite{Balzer2024}. dKMC is a coarse-grained, effective-Hamiltonian method that requires a lattice of molecular sites, parameterised by their energies, couplings, spectral densities, and recombination rates. Sec.~S1 of the SI includes a summary of the parameters used. Here, we focus on the atomistic simulations of Y6 that we use to parameterise the dKMC simulations of Y6.

\subsection{\label{subsec:atomistic}Atomistic Calculations}

\textbf{General details:} Parameters for dKMC are calculated using density functional theory~(DFT), using the long-range-corrected CAM-B3LYP exchange–correlation functional with the 6–31+G(d,p) basis set~\cite{Yanai2004}. The exchange–correlation functional is range-tuned according to the established non-empirical procedure, resulting in an optimal range-separation parameter of \SI{0.015}{\bohr^{-1}}~\cite{Kronik2012,Sun2016,Refaely2013,Chen2018}. Linear response TD-DFT calculations for excited states are performed within the Tamm–Dankoff approximation~(TDA), with dielectric stabilisation treated using a polarisable continuum model with static dielectric constant set to~3.8.
Single-molecule geometry optimisations (for reorganisation-energy and normal-mode calculations) are performed in vacuum, with the standard (non-range tuned) CAM-B3LYP functional, and with alkyl chains truncated to ethyl. The optimised molecular geometries are confirmed by calculation of the associated vibrational modes, which show no imaginary frequencies.
All \mbox{(TD-)DFT} calculations and geometry optimisations are performed using Gaussian~16, Rev.~C.01~\cite{Gaussian}. 

\textbf{Reorganisation energies:} For classical KMC, reorganisation energies $\lambda_\mathrm{i \xrightarrow{}f}$ for transfer from initial electronic state `i' to final electronic state `f' are calculated using the four-point scheme~\cite{Bredas2004}, $\lambda_\mathrm{i \xrightarrow{}f} = (E_\mathrm{1,i}(\textbf{R}_\mathrm{1,f}) -E_\mathrm{1,i}(\textbf{R}_\mathrm{1,i} )) + (E_\mathrm{2,i}( \textbf{R}_\mathrm{2,f} )-E_\mathrm{2,i}( \textbf{R}_\mathrm{2,i} ))$, where, for example, $E_\mathrm{1,i}(\textbf{R}_\mathrm{1,f})$ is the energy of molecule~1 in state~`i' in the optimised geometry of the state~`f'.

\textbf{Diabatic state energies and couplings:} 
Energies and couplings for exciton and CT states (and couplings of these states to the ground state) are calculated on pairs of molecules extracted from the X-ray crystal structure~\cite{Zhang2020}. First, reference calculations are performed on pairs of molecules separated by approximately \SI{20}{\angstrom}. At this distance, negligible orbital overlap results in localised CT states and pure exciton states that can be localised by diabatization~\cite{Hsu2008}. Energies and couplings for other molecular pairs (with smaller separations) are then calculated by projecting the reference states onto the states of the molecular pairs in the crystal geometry~\cite{Hume2020,Hume2021,Price2022}. 
The CT-state energies calculated in this way are characteristic for a given pair. However, the energies of the localised exciton states differ slightly depending on which pair is used to calculate them. Therefore, we take the energy of an exciton on a given molecule to be the mean of the values obtained from the relevant pair calculations.
Electronic couplings for ground-state electron and hole transfer are calculated using the dimer projection (DIPRO) method~\cite{Baumeier2010,Valeev2006}. Importantly, all couplings are calculated using a common set of reference geometries, so that each molecule in a dimer can be related to one of the reference geometries by a pure translation. Doing so ensures that all couplings are calculated using a consistent orbital phase.  

\subsection{Parameterising dKMC}

\textbf{Sites:} dKMC models transport processes on a lattice of molecular sites, each positioned in the centre of a Y6 molecule in its crystal structure. The lattice consists of $(4N)^d$ sites, formed by replicating $N$ unit cells (each with two pairs of equivalent molecules) along each of the $d$ dimensions, \cref{fig:model}. Although our lattices are always set up in three-dimensional space, a 1D simulation involves replicating the unit cell in the direction of only one primitive lattice vector, while a 2D simulation involves replicating it along two primitive lattice vectors.

\textbf{One-particle Hamiltonian:}
The Hilbert space of the one-particle transport problem is spanned by sites $\ket{n}$ representing an electron, hole, or exciton on site $n$. The system Hamiltonian in this basis is
\begin{equation}
    H_\mathrm{S} = \sum_{n} E^z_m \ket{n}\bra{n} + \sum_{n'\neq n} J^z_{nn'} \ket{n}\bra{n'}, 
\end{equation}
where $z$ is a superscript indicating the particle type (`e', `h', or `xc'), $E^z_n$ is the energy of site $n$, and $J^z_{nn'}$ is the coupling between sites $n$ and $n'$. Parameterisations of energies and couplings are given below.

\textbf{Two-particle Hamiltonian:}
The Hilbert space of the two-particle charge-generation problem is spanned by site-pairs $\ket{m,n}$ representing the state of an electron on site $m$ and a hole on site $n$. The two-particle Hamiltonian is then
\begin{multline}
    H_\mathrm{S} = \sum_{m,n} E_{mn} \ket{m,n}\bra{m,n} + \!\!\! \sum_{m\neq m', n} \!\!\! J^\mathrm{e}_{mm'} \ket{m,n}\bra{m',n} \\ 
    + \!\! \sum_{m, n\neq n'} \!\!\! J^\mathrm{h}_{nn'} \ket{m,n}\bra{m,n'} + \sum_{o \neq o'} J^\mathrm{xc}_{oo'} \ket{o,o}\bra{o',o'}
    \label{eq:H_S},
\end{multline}
where site-pair energies are $E_{mn}=E_m^\mathrm{LUMO}-E_n^\mathrm{HOMO} + U(r_{mn})$, with $U(r_{mn})$ the Coulomb interaction. Two site-pairs are coupled when only one of the carriers is moving. In the first term, two site-pairs are coupled with the electron coupling $J^\mathrm{e}_{mm'}$ when the hole is stationary ($n=n'$) and the electron moves between sites $m$ and $m'$ (and analogously for the hole coupling $J^\mathrm{h}_{nn'}$). Finally, two site-pairs are coupled with the exciton coupling $J^\mathrm{xc}_{oo'}$ when both describe an exciton ($m=n=o$ and $m'=n'=o'$) that moves from site $o$ to site $o'$. The parameterisations of the site-pair energies and couplings are also described below.

\textbf{Site energies:} To describe disorder in realistic materials, each site is assigned HOMO and LUMO energies, drawn from disordered energy distributions (\cref{fig:model}b).

The mean energies $E_M^\mathrm{LUMO}$ and $E_M^\mathrm{HOMO}$ for each of the four molecules in the unit cell ($M\in \{1,2,3,4\}$) are determined from the CT-state energies of pairs of molecules separated by \SI{300}{\angstrom}. At this distance, the Coulomb interaction is invariant to interchange of the electron and hole positions. Performing these calculations for all 16 types of molecular pairs results in a system of simultaneous equations that are then solved for $E_M^\mathrm{LUMO}$ and $E_M^\mathrm{HOMO}$. Using dimer calculations to determine local energies ensures that consistent electronic-structure techniques are used for both one-particle and two-particle Hamiltonian parameters.

Energetic disorder is modelled by Gaussian distributions $\mathcal{N}(E_0,\sigma)$ with mean $E_0$ and standard deviation $\sigma$~\cite{Bassler1993}. The HOMO and LUMO energies of molecule $M$ in unit cell $C$ are drawn from the bivariate normal distribution
\begin{equation}
\label{eq:energy_distribution}
\begin{bmatrix}
E_{M,C}^\mathrm{LUMO} \\ 
E_{M,C}^\mathrm{HOMO}
\end{bmatrix}
\sim \mathcal{N}\left(
\begin{bmatrix}
E_M^\mathrm{LUMO}\\ 
E_M^\mathrm{HOMO}
\end{bmatrix},
\begin{bmatrix}
\sigma_\mathrm{e}^2 & \rho\sigma_\mathrm{e}\sigma_\mathrm{h} \\ 
\rho\sigma_\mathrm{e}\sigma_\mathrm{h} & \sigma_\mathrm{h}^2
\end{bmatrix}
\right).
\end{equation}
Therefore, the HOMO and LUMO energies have disorders $\sigma_\mathrm{h}$ and $\sigma_\mathrm{e}$, respectively, while their correlation coefficient $\rho = \left(\sigma_\mathrm{e}^2 + \sigma_\mathrm{h}^2 + \sigma_\mathrm{xc}^2\right)/\left(2\sigma_\mathrm{e}\sigma_\mathrm{h}\right)$ allows the excitonic disorder $\sigma_\mathrm{xc}$ to be smaller than the energetic disorders. The energetic disorders $\sigma_{e}=\SI{100}{meV}$ and $\sigma_\mathrm{h}=\SI{103}{meV}$ are calculated by fitting Gaussian profiles to the distributions of electron and hole energies of a simulated Y6 film taken from previous work~\cite{Price2022}, where the Y6 crystal structure was used to initiate a molecular-dynamics simulation at \SI{300}{K}. Snapshots from this simulation were used to perform electrostatic-embedding calculations of the electron and hole energies associated with different molecular sites~\cite{Poelking2015,Poelking2015_2,Poelking2016}. The exciton disorder $\sigma_\mathrm{xc}=\SI{56}{meV}$ is obtained from temperature-dependent photoluminescence measurements, specifically the slope of the 0--0 peak position with inverse temperature~\cite{Firdaus2020}.

\textbf{Coulomb interactions:}
The Coulomb interaction $U(r_{mn})$ is also parameterised using atomistic calculations. For pairs of molecules with any non-hydrogen atoms less than \SI{8}{\angstrom} apart, DFT is used to calculate the total site-pair energy $E_{mn}$, accounting for near-field electron-hole interactions that may not be captured by a simple Coulomb law. 
For site-pairs with larger separations, the Coulomb interaction is included via the Coulomb law $U(r_{mn})\propto 1/r_{mn}$, where the proportionality constant is calculated from fits of the energies of pairs of molecules in the Y6 crystal at various separations (see sec.~S2 of the SI). 

\textbf{Couplings:}
All coupling strengths are calculated using the methods described in \cref{subsec:atomistic}. Couplings $J^\mathrm{e}_{nn'}$, $J^\mathrm{h}_{nn'}$, and $J^\mathrm{xc}_{nn'}$ are calculated for pairs of molecules that are close enough to be significantly coupled, which we take to be given by the same \SI{8}{\angstrom} cutoff as for the Coulomb interaction above. For pairs of molecules with greater separations, couplings are assumed to be zero. In the two-particle Hamiltonian, the electron, hole, and exciton couplings are calculated in the same way, except when charge transfer involves movement to or from an exciton site-pair; those couplings are calculated explicitly to account for complicated electron-hole interactions.

\textbf{Spectral densities:}
Spectral densities describe the interaction of the system with its environment (or bath) at different frequencies. In dKMC, we assume an independent bath of harmonic oscillators on each site, $H_\mathrm{B}=\sum_{m,k}\omega_{mk} b^\dag_{mk}b_{mk}$, with the $k$th mode on site $m$ having frequency $\omega_{mk}$ and annihilation operator $b_{mk}$. The system-bath interaction Hamiltonian
\begin{multline}
     \label{eq:H_SB}
     H_\mathrm{SB} = \sum_{m,n\neq m,k} g^\mathrm{e}_{mk}\ket{m,n}\bra{m,n}(b^\dag_{mk} + b_{mk})\\
    + \sum_{m,n\neq m,k} g^\mathrm{h}_{nk}\ket{m,n}\bra{m,n}(b^\dag_{nk} + b_{nk})\\
    +\sum_{o,k} g^\mathrm{xc}_{ok}\ket{o,o}\bra{o,o}(b^\dag_{ok} + b_{ok}).
\end{multline}
describes a linear coupling of the site-pair energies to the bath modes with strength $g^\mathrm{e}_{mk}$ for electrons, $g^\mathrm{h}_{nk}$ for holes, and $g^\mathrm{xc}_{ok}$ for excitons. The interaction is simplified by assuming that the couplings are equal for all sites ($g^z_{nk}=g^z_k$) and by replacing the discrete spectral density $j^z(\omega)=\sum_k \left(g^z_k\right)^2\delta(\omega-\omega_k)$ with a continuous one.

To calculate the spectral densities for Y6, we start by calculating the normal-mode frequencies ($\omega_k$) of Y6 and their Huang-Rhys factors ($S^\mathrm{e}_k$ for electrons, $S^\mathrm{h}_k$ for holes, and $S^\mathrm{xc}_k$ for excitons). These are calculated from the optimised molecular geometries using the Franck-Condon method in Gaussian 16. Huang-Rhys factors give the system-bath couplings $g^z_k = \sqrt{S^z_k}$. 

In principle, these normal modes and Huang-Rhys factors could be used to parameterise the spectral density of Y6. However, many of the normal modes of Y6 are much slower than the transport timescale---i.e., their frequency is much less than the inverse of a typical transfer rate---violating the assumption of fully displaced bath modes implicit in the full polaron transformation used in dKMC. The polaron transformation~\cite{Grover1971}
\begin{multline}
     e^S = \exp\Bigg(\sum_{m,n\neq m,k} \frac{g^\mathrm{e}_{mk}}{\omega_{mk}}\ket{m,n}\bra{m,n}(b^\dag_{mk}-b_{mk}) \\
     + \sum_{m,n\neq m,k}\frac{g^\mathrm{h}_{nk}}{\omega_{nk}}\ket{m,n}\bra{m,n}(b^\dag_{nk}-b_{nk}) \\
     +
     \sum_{o,k}\frac{g^\mathrm{xc}_{ok}}{\omega_{ok}}\ket{o,o}\bra{o,o}(b^\dag_{ok}-b_{ok})
     \Bigg),
     \label{eq:polaron}
\end{multline}
when applied to the entire Hamiltonian ($H = H_\mathrm{S} + H_\mathrm{B} + H_\mathrm{SB}$), displaces the bath modes to describe the formation of polarons, quasi-particles that contain carriers and the distortion they induce in the environment~\cite{Frolich1954,Holstein1959}.

Slow bath modes do not have time to fully displace before the carrier moves on from a site. A more accurate way to deal with them would be the variational polaron transformation~\cite{Pollock2013,Jang2022}, which calculates partial mode displacements by minimising the upper
bound of an effective free energy (using the Feynman-Bogoliubov inequality) of the system. As a result, it predicts, for every mode, the variational parameter $f^z_k$ that replaces $g^z_k$ and describes the extent of the mode's displacement. After the variational polaron transformation, modes that are much faster than typical transport timescales are fully displaced ($f^z_k\approx g^z_k$), those that are much slower are not displaced at all ($f^z_k\approx 0$), while those in between are partially displaced ($f^z_k < g^z_k$). However, the variational polaron transformation requires large, self-consistent minimisations of the effective free energy of the entire system, preventing its use in large simulations like ours.

Instead, we take an intermediate approach inspired by the frozen-mode small-polaron quantum master equation (SPQME)~\cite{Teh2019}, which avoids the computational complexity of the variational treatment while reducing the error associated with slow baths. The frozen-mode approach uses the variational polaron framework, but with two simplifications. First, the variational treatment is used only for a representative dimer, which has coupling $J_0$ and energy difference $\Delta$. Second, this dimer is used to calculate a splitting frequency that divides slow bath modes that are frozen (and treated classically) from fast ones that are fully displaced.

We use a similar approach, adopting the first approximation but modifying the second to improve accuracy. Instead of assuming a sharp transition at a single splitting frequency, we find the displacement of every bath mode in the representative dimer. We choose our representative dimer to have coupling $J^z_0=\mathrm{max}(J^z)$ and energy difference $\Delta=0$, as these represent the fastest transfers, which are most important for the polaron transformation to calculate correctly. Next, we choose our initial guess for $f^z_k(\omega_k)/g^z_k(\omega_k)$ to be a linear ramp from 0 to 1 across the range of bath-mode frequencies. Then, we variationally optimise $f^z_k(\omega_k)/g^z_k(\omega_k)$ by iteratively applying the formulas
\begin{align}
    \frac{f^z_k}{g^z_k} &=\bigg(1 + \frac{\tanh\big(\beta\sqrt{\Delta^2+J_\mathrm{eff}^2}\big)J_\mathrm{eff}^2\coth{(\beta\omega_k/2)}}{\omega_k\sqrt{\Delta^2+J_\mathrm{eff}^2}}\bigg)^{-1},\nonumber\\
    J_\mathrm{eff} &= J^z_0\exp{\bigg(-\sum_k\frac{(f^z_k)^2}{\omega_k^2}\coth{\left(\beta\omega_k/2\right)}\bigg)},
\end{align}
until subsequent iterations of the $f^z_k(\omega_k)/g^z_k(\omega_k)$ vector differ by less than 1\% in root-mean-squared distance. Doing so minimises the effective free energy of the representative dimer, and gives the variational system-bath couplings $f^z_k$ that we use for every bath mode (detailed in sec.~S3 of the SI). 

As a result, our discrete spectral density is $j^z(\omega)=\sum_k \left(f^z_k\right)^2\delta(\omega-\omega_k)$, with variationally determined couplings $f^z_k$ instead of $g^z_k$. Finally, we convert the discrete spectral density into a continuous one by convolving each line with a Lorentzian lineshape of width $\qty{e12}{s^{-1}}$, corresponding to typical vibrational dephasing rates~\cite{Nitzan2024}. The choice of line width is not particularly significant, except for determining the limits of numerical integration when calculating the bath correlation terms. Sec.~S3 of the SI includes plots of each spectral density with and without the variational corrections.

\textbf{Recombination rates:}
The recombination of excitons and CT states is treated as described in previous dKMC work~\cite{Balzer2022,Balzer2024}. While classical KMC typically uses constant recombination rates ($R_\mathrm{recomb}^\mathrm{xc}$ and $R_\mathrm{recomb}^\mathrm{CT}$), dKMC uses a rate derived from Fermi's golden rule that corresponds to the same recombination rate modified by a delocalisation correction to reflect that delocalised states are longer lived because they have lower overlap with local excitations that decay rapidly. In particular, the exciton recombination rate of state $\nu$ is~\cite{Balzer2024}
\begin{equation}
    \label{eq:xc_recomb}
    k^\nu_\mathrm{xc,recomb}= R_\mathrm{recomb}^\mathrm{xc}\bigg|\sum_{o}\braket{\nu|o,o}\bigg|^2.
\end{equation}
We use $R_\mathrm{recomb}^\mathrm{xc}=1/L^\mathrm{xc}$, the inverse of the measured exciton lifetime of dilute Y6 in polystyrene film, $L^\mathrm{xc} = \SI{260}{ps}$~\cite{Firdaus2020}. 

Applying the same approach to CT-state recombination is difficult for two reasons. First, CT-state lifetimes are not known precisely because CT states in Y6 are close in energy (and coupled) to exciton states, making it difficult to distinguish them experimentally. Second, it is unclear which site-pairs in the Y6 crystal should be considered CT site-pairs. Instead, we assume that the fastest CT site-pair recombination---from the site-pair with the strongest coupling to the ground state, $\mathrm{max}(J^g)$---occurs at the same rate as exciton recombination. This choice is conservative because it overestimates the rate of CT recombination, meaning that IQEs will be at least as large as calculated. Then, all other site-pairs recombine with the rate scaled down by the factor $\abs{J_{mn}^g/\mathrm{max}(J^g)}^2$, which reflects their weaker coupling to the ground state. Finally, the CT recombination rate of any state $\nu$ is therefore
\begin{equation}
     \label{eq:CT_recomb}
     k^\nu_\mathrm{CT,recomb}= R_\mathrm{recomb}^\mathrm{xc}\Bigg|\frac{\sum_{m,n}\braket{\nu|m,n}J_{mn}^g}{\mathrm{max}(J^g)}\Bigg|^2.
\end{equation}

\subsection{dKMC dynamics}
In dKMC, dynamics proceeds by hopping through partially delocalised polaronic states, which are the eigenstates of the polaron-transformed system Hamiltonian $\tilde{H}_\mathrm{S}$. By applying the polaron transformation to the entire Hamiltonian ($H = H_\mathrm{S} + H_\mathrm{B} + H_\mathrm{SB}$), most of the system-bath interaction is moved into the formation of polarons, meaning that the states are less delocalised in the polaron frame due to the renormalisation of the electronic coupling~\cite{Rice2018}. We quantify the extent of delocalisation with the inverse participation ratio, which, for a polaron state $\nu$ is 
\begin{equation}
    \label{eq:IPR}
    \mathrm{IPR}_\nu = \Big(\sum_n \abs{\braket{n|\nu}}^4\Big)^{-1}.
\end{equation}
The remaining weak system-bath interaction can then be treated perturbatively, which, when done to second order, leads to the secular polaron-transformed Redfield master equation (sPTRE)~\cite{Lee2015}, the quantum master equation underpinning dKMC. dKMC stochastically unravels the sPTRE, converting it into a kinetic Monte Carlo procedure that tracks and averages over many stochastic trajectories through the polaron states. Below, we outline the dKMC procedure, whose details were described previously for both one-particle~\cite{Balzer2020,Balzer2023} and two-particle~\cite{Balzer2022,Balzer2024} simulations.

\textbf{Transport:}
dKMC simulations begin with a disordered energy landscape of the Y6 lattice sites. To choose an initial state close to the centre of the simulation box, we first select a random site from the unit cell in the middle of the lattice. Next, we diagonalise a subset of $\tilde{H}_\mathrm{S}$ that contains sites within the neighbourhood of the chosen site. The size of this neighbourhood is defined by the Hamiltonian radius $r_\mathrm{ham}$, which is calibrated to be large enough to contain the current state and the states to which it is likely to move~\cite{Balzer2023,Balzer2020}. Finally, we choose the initial state as the one with the greatest overlap with the chosen site.

Next, we propagate the dynamics stochastically. At each step, we calculate the hopping rates from the current state to other states that are within a cut-off distance called the hopping radius, $r_\mathrm{ham}$. The hopping radius is also precalculated and calibrated to capture the total outgoing hopping rates to a desired accuracy, $a_\mathrm{dKMC}$. The next state is chosen from among the possible destinations probabalistically, in proportion to the magnitude of the outgoing hopping rates, before updating the elapsed time and rediagonalising a new subset of $\tilde{H}_\mathrm{S}$ centred at the new state. This procedure continues for a simulation time of $t_\mathrm{end}=\SI{100}{ps}$, which corresponds to transit times on typical length scales (tens of nanometres) in organic semiconductors.

The full procedure above is repeated for $n_\mathrm{traj}$ trajectories on $n_\mathrm{iters}$ energetic landscapes, before calculating the mean-squared displacement as a function of time, $\overline{\langle r^2(t)\rangle}$. For exciton transport, we use the mean-squared displacement to calculate the diffusion coefficient
\begin{equation}
    \label{eq:D}
    D=\lim_{t\to\infty}\frac{d}{dt}\left(\frac{\overline{\langle r^2(t)\rangle}}{6}\right),    
    \end{equation}
while for electron and hole transport we convert the diffusion coefficient to a mobility via the Einstein relation
\begin{equation}
    \label{eq:einstein_mobility}
    \mu=\frac{eD}{k_BT},
\end{equation}
where $e$ is the elementary charge, $k_B$ is the Boltzmann constant, and $T$ is the temperature.

\textbf{Charge generation:}
Charge generation simulations begin with the same disordered energy landscape. Next, also as above, we choose the initial state to be an exciton state close to the middle of the system. We then propagate the dynamics stochastically in the same way as for transport, except that possible transitions at each step include electron, hole, and exciton transfer, as well as recombination through overlap with exciton or CT site-pairs, as described in \cref{eq:xc_recomb,eq:CT_recomb}.

This procedure continues until the charges separate, which we consider to be an electron-hole separation greater than $r_\mathrm{sep}=\SI{5}{nm}$, or they recombine. Charges are also considered recombined if the number of hops exceeds a maximum, $n_\mathrm{hops}=10000$, which is included to prevent infinite loops, such as those that happen when charges get caught between two energetic traps. The complete procedure from initialisation to termination is repeated for $n_\mathrm{traj}$ trajectories on $n_\mathrm{iters}$ energetic landscapes, before calculating the IQE as the percentage of all trajectories where charges separate.

\begin{figure*}
    \centering
    \includegraphics[width=\textwidth]{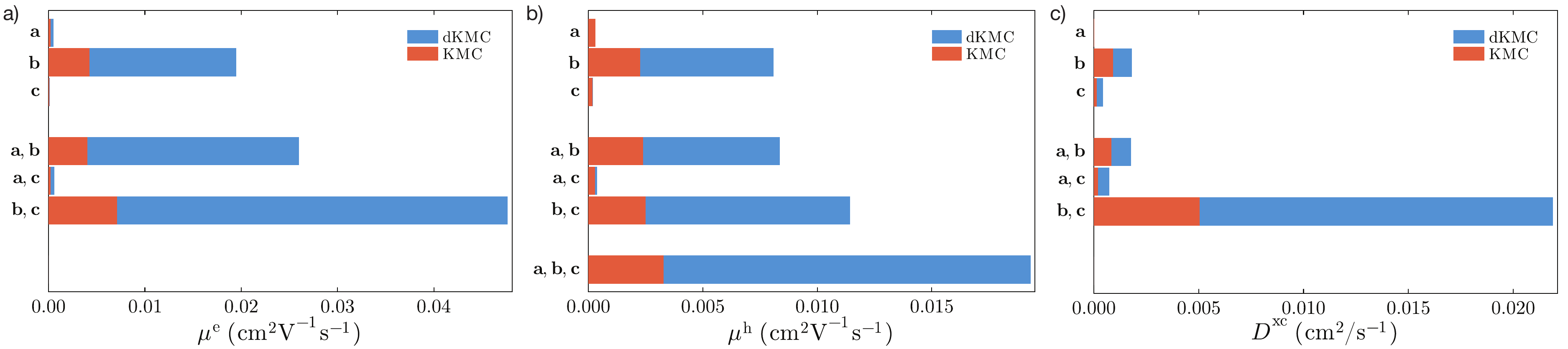}
    \caption{\textbf{Transport properties of Y6} calculated using classical KMC and dKMC. \textbf{(a)}~Electron mobilities, \textbf{(b)}~hole mobilities, and \textbf{(c)}~exciton diffusion coefficients. Included are results for lattices with different dimensionality where the unit cell is replicated in the direction of one or more of the primitive lattice vectors \vec{a}, \vec{b}, and \vec{c}. In all cases, when delocalisation is included using dKMC, the transport is significantly faster compared to classical KMC results.
    }
    \label{fig:mobilities}
\end{figure*}

\begin{table}
	\centering
    \begin{tabular}{lllll}
		\toprule
		1D & $\mu^\mathrm{e}_\mathrm{KMC}~(\mathrm{cm^2/Vs})$ & $\mu^\mathrm{e}_\mathrm{dKMC}~(\mathrm{cm^2/Vs})$ & $\dfrac{\mu^\mathrm{e}_\mathrm{dKMC}}{\mu^\mathrm{e}_\mathrm{KMC}}$ & $\overline{\mathrm{IPR}^\mathrm{e}}$ \\
		\midrule
		\vec{a} & \num{1.9 \pm 0.2 e-4} & \num{5.1 \pm 0.3 e-4} & \num{2.6 \pm 0.3} & \num{1.419 \pm 0.009}\\
		\vec{b} & \num{4.25 \pm 0.19 e-3} & \num{1.84 \pm 0.06 e-2} & \num{4.3 \pm 0.2} & \num{2.005 \pm 0.017} \\
		\vec{c} & \num{8.0 \pm 1.4 e-5} & \num{1.22 \pm 0.14 e-4} & \num{1.5 \pm 0.3} & \num{1.355 \pm 0.008}\\
        \midrule
        2D &  &  & \\
        \midrule
        \vec{a},\vec{b} & \num{4.02 \pm 0.14 e-3} & \num{2.57 \pm 0.07 e-2} & \num{6.4 \pm 0.3} & \num{2.58 \pm 0.03}\\
        \vec{a},\vec{c} & \num{1.94 \pm 0.15 e-4} & \num{5.8 \pm 0.3 e-4} & \num{3.0 \pm 0.3} & \num{1.442 \pm 0.009}\\
        \vec{b},\vec{c} & \num{7.1 \pm 0.2 e-3} & \num{4.73 \pm 0.10 e-2} & \num{6.7 \pm 0.2} & \num{2.75 \pm 0.03}\\
        \midrule
        3D & & & \\
        \midrule
        \vec{a},\vec{b},\vec{c} & \num{9.1 \pm 0.3 e-3} &  &  & \\
        \bottomrule
        & \\
        \toprule
        1D & $\mu^\mathrm{h}_\mathrm{KMC}~(\mathrm{cm^2/Vs})$ & $\mu^\mathrm{h}_\mathrm{dKMC}~(\mathrm{cm^2/Vs})$ & $\dfrac{\mu^\mathrm{h}_\mathrm{dKMC}}{\mu^\mathrm{h}_\mathrm{KMC}}$ & $\overline{\mathrm{IPR}^\mathrm{h}}$\\
		\midrule
		\vec{a} & \num{3.1 \pm 0.2 e-4} & \num{2.29 \pm 0.15 e-4} & \num{0.73 \pm 0.07} & \num{1.136 \pm 0.005} \\
		\vec{b} & \num{2.27 \pm 0.12 e-3} & \num{8.8 \pm 0.3 e-3} & \num{3.9 \pm 0.3} & \num{1.591 \pm 0.011} \\
		\vec{c} & \num{1.83 \pm 0.18 e-4} & \num{1.97 \pm 0.15 e-4} & \num{1.08 \pm 0.13} & \num{1.142 \pm 0.006}\\
        \midrule
        2D &  &  & \\
        \midrule
        \vec{a},\vec{b} & \num{2.39 \pm 0.09 e-3} & \num{9.3 \pm 0.3 e-3} & \num{3.9 \pm 0.2} & \num{1.580 \pm 0.013}\\
        \vec{a},\vec{c} & \num{2.93 \pm 0.18 e-4} & \num{3.5 \pm 0.2 e-4} & \num{1.20 \pm 0.10} & \num{1.139 \pm 0.005}\\
        \vec{b},\vec{c} & \num{2.51 \pm 0.10 e-3} & \num{1.12 \pm 0.03 e-2} & \num{4.5 \pm 0.2} & \num{1.785 \pm 0.014}\\
        \midrule
        3D & & & \\
        \midrule
        \vec{a},\vec{b},\vec{c} & \num{3.28 \pm 0.12 e-3} & \num{1.96 \pm 0.05 e-2} & \num{6.0\pm 0.3} & \num{2.073 \pm 0.019}\\
        \bottomrule
        & \\
        \toprule
        1D & $D^\mathrm{xc}_\mathrm{KMC}~(\mathrm{cm^2/s})$ & $D^\mathrm{xc}_\mathrm{dKMC}~(\mathrm{cm^2/s})$ & $\dfrac{D^\mathrm{xc}_\mathrm{dKMC}}{D^\mathrm{xc}_\mathrm{KMC}}$ & $\overline{\mathrm{IPR}^\mathrm{xc}}$\\
        \midrule
        \vec{a} & \num{3.5 \pm 0.3 e-6} & \num{7.9 \pm 0.3 e-6} & \num{2.2 \pm 0.2} & \num{1.148 \pm 0.004}\\
        \vec{b} & \num{9.23 \pm 0.15 e-4} & \num{1.88 \pm 0.03 e-3} & \num{2.04 \pm 0.04} & \num{1.302 \pm 0.007}\\
        \vec{c} & \num{1.37 \pm 0.03 e-4} & \num{4.42 \pm 0.08 e-4} & \num{3.24 \pm 0.10} & \num{1.210 \pm 0.005}\\
        \midrule
        2D &  &  & \\
        \midrule
        \vec{a},\vec{b} & \num{8.30 \pm 0.13 e-4} & \num{1.76 \pm 0.03 e-3} & \num{2.12 \pm 0.05} & \num{1.441 \pm 0.010}\\
        \vec{a},\vec{c} & \num{2.02 \pm 0.04 e-4} & \num{7.56 \pm 0.14 e-4} & \num{3.74 \pm 0.10} & \num{1.346 \pm 0.008}\\
        \vec{b},\vec{c} & \num{5.05 \pm 0.04 e-3} & \num{2.23 \pm 0.03 e-2} & \num{4.42 \pm 0.06} & \num{2.026 \pm 0.017}\\
        \midrule
        3D & & & \\
        \midrule
        \vec{a},\vec{b},\vec{c} & \num{4.90 \pm 0.07 e-3} &  &  & \\
        \bottomrule
    \end{tabular}    
    \caption{
	   \textbf{Transport properties of Y6} calculated using classical KMC and dKMC. \textbf{Top:}  electron mobility $\mu^\mathrm{e}$; \textbf{middle:} hole mobility $\mu^\mathrm{h}$; \textbf{ bottom:} exciton diffusion coefficient $D^\mathrm{xc}$. For all three transport properties, delocalisation---accounted for in dKMC---significantly improves transport, as quantified by the delocalisation enhancement $\mu_\mathrm{dKMC}^{z}/\mu_\mathrm{KMC}^{z}$. The extent of delocalisation of each carrier is quantified by the mean IPR of all states, $\overline{\mathrm{IPR}^z}$. Each row represents a lattice with different dimensionality, where the unit cell is replicated in the direction of one or more of the primitive lattice vectors \vec{a}, \vec{b}, or \vec{c}. 3D results are not given for electrons or excitons as the calculations were intractable. The uncertainties come from the standard deviations of the linear fits of \cref{eq:D}.
    }
	\label{tab:transport}
\end{table}

\section{Results}

\subsection{Transport}

\Cref{tab:transport} and \cref{fig:mobilities} display the results of electron, hole, and exciton transport simulations in Y6, calculated using both dKMC, which accounts for delocalisation, and classical KMC, which does not. Classical KMC assumes that charges are localised onto individual sites and move via thermally assisted hops, modelled as Marcus transfer~\cite{Marcus1956}. \Cref{tab:transport} includes results for lattices with different dimensionality, in which the Y6 unit cell has been replicated in the direction of one or more of the primitive lattice vectors, because our previous work has shown the importance of considering dimensionality when modelling transport~\cite{Balzer2020,Balzer2022,Balzer2023,Balzer2024}. For 1D simulations, it is replicated in the direction of one of the primitive lattice vectors (\vec{a}, \vec{b}, or \vec{c}); in 2D simulations, along two of them (\vec{a} and \vec{b}, \vec{a} and \vec{c}, or \vec{b} and \vec{c}), and in 3D, along all three. However, 3D calculations can become intractable for dKMC due to the need to diagonalise large Hamiltonians; therefore, we only report hole mobilities in 3D because holes are less delocalised than electrons or excitons.

Our results show that delocalisation improves the transport of all three carriers in Y6 and in all dimensions. The magnitude of the improvement is quantified by the delocalisation enhancement $\mu_\mathrm{dKMC}^{z}/\mu_\mathrm{KMC}^{z}$, which we find to be as much as a factor of \num{6.7 \pm 0.2} for electrons, \num{6.0 \pm 0.3} for holes, and \num{4.42 \pm 0.06} for excitons. The results highlight the importance of modelling delocalised transport in higher dimensions, even for highly anisotropic materials such as Y6, as delocalisation (measured by IPR) increases markedly with dimensionality. For all carriers, transport along the \vec{b} direction, which corresponds to the direction of $\pi$--$\pi$ stacking, is significantly faster than in either the \vec{a} or \vec{c} directions. However, for all carriers, the transport and delocalisation enhancements significantly improve when the lattice is extended in the direction of a second primitive lattice vector. For holes, replicating the lattice in all three directions increases the hole mobility and the delocalisation enhancement by an additional factor of 1.75 above the highest 2D value. While these increases with dimensionality are also seen classically due to the increased numbers of possible transport pathways, the effect is much greater for dKMC than for KMC, because of the additional benefit of increased delocalisation with dimensionality. Therefore, we expect that the electron mobilities, exciton diffusion coefficients, and mean IPRs would be even higher in 3D than their maximum values in 2D. For a rough estimate, the 1.75-fold enhancement in hole mobilities going from 2D to 3D could be used to extrapolate the 2D electron mobilities and exciton diffusion coefficients to 3D. However, to remain conservative in our claims about the role of delocalisation, we use the maximum calculated 2D values below.

Comparing theoretical transport predictions with experimental results is difficult in organic semiconductors because many experimental factors are difficult to control, leading to a wide range of reported measurements. The factors that are most difficult to control are the morphology (including crystallinity and grain boundaries) and purity of the material (including traps and adventitious doping). Transport properties also depend on the temperature, the electric field, and the carrier density, which can vary from one experiment to another. 

Comparison is easier for excitons, where including delocalisation yields agreement between theoretical and experimental diffusion coefficients. Experimental diffusion coefficients in Y6 tend to fall in the range of \numrange{1.7e-2}{3.6e-2}~$\mathrm{cm}^2\mathrm{s}^{-1}$~\cite{Firdaus2020,Natsuda2021,LoGerfo2023}, although it should be noted that these values assume a purely excitonic model that neglects charge-pair formation. Classical KMC predictions fall well below this range, with a maximum of \num{5.05\pm0.04 e-3}~$\mathrm{cm}^2\mathrm{s}^{-1}$. However, when delocalisation is included with dKMC, the predicted diffusion coefficients increase into the experimental range, with the largest 2D value of \num{2.23\pm0.03 e-2}~$\mathrm{cm}^2\mathrm{s}^{-1}$.

Comparison is more difficult for charge carriers because charge transport is a time-dependent phenomenon. Organic semiconductors are often so disordered that transport does not reach equilibrium values on realistic timescales~\cite{Hoffmann2012,Melianas2019}, meaning that the mobility is a function of the timescale used. For this reason, different experimental techniques can return significantly different mobilities for the same material. For example, time-of-flight (TOF) and space-charge-limited current (SCLC) measure long-time, device-level mobilities, which are much lower than the mobilities obtained by optical-pump tetrahertz-probe (OPTP) spectroscopy and time-resolved microwave conductivity (TRMC), which probe short-time, local mobilities. In addition, the longer-scale experiments are more sensitive to grain boundaries and traps than the shorter-scale ones.

Despite the more difficult comparison, delocalisation also improves the agreement between theoretical and experimental charge-carrier mobilities. The range of reported mobilities is much larger than that for exciton diffusion coefficients; for example, reported electron mobilities in Y6 span four orders of magnitude, \num{0.00065}--\SI{2.4}{cm^2V^{-1}s^{-1}}~\cite{Yao2021,Saglamkaya2023,Firdaus2020,Gutierrez2022}. However, the extremes of this range are results of measurements that are not comparable with our simulations. The high end of the range (\SI{2.4}{cm^2V^{-1}s^{-1}}) comes from measurements of single crystals~\cite{Gutierrez2022}, whose disorder is lower than considered here. By contrast, the low end comes from long-time device measurements~\cite{Yao2021,Saglamkaya2023}, which are dominated by traps, grain boundaries, and other device-scale impurities not included in our calculations. 
The range is narrower for techniques that correspond more closely to the local, short-time mobilities calculated in this work. 
These include TRMC and OPTP, but they can only report $\varphi\Sigma\mu=\varphi(\mu^\mathrm{e}+\mu^\mathrm{h})$, the product of the IQE $\varphi$ with the sum of the electron and hole mobilities. TRMC measurements on Y6 gave $\varphi\Sigma\mu_\mathrm{TRMC}=\SI{0.0095}{cm^2V^{-1}s^{-1}}$~\cite{Firdaus2020}, while OPTP measurements $\varphi\Sigma\mu_\mathrm{OPTP}=0.01-\SI{0.1}{cm^2V^{-1}s^{-1}}$~\cite{Price2022,Li2024}. While this makes the direct comparison of either electron or hole mobilities difficult, it can be used for a combined comparison of both mobilities and IQE. Using our calculated IQEs, discussed in \cref{subsec:IQE}, classical KMC under-predicts the TRMC/OPTP measurements by over an order of magnitude even in 3D, $\varphi\Sigma\mu_\mathrm{KMC}=\SI{0.00068\pm0.00002}{cm^2V^{-1}s^{-1}}$. When delocalisation is included, the agreement improves significantly, with dKMC predictions lying within the experimental range, $\varphi\Sigma\mu_\mathrm{dKMC}=\SI{0.0139\pm0.0004}{cm^2V^{-1}s^{-1}}$.

\subsection{Charge generation\label{subsec:IQE}}

\begin{figure}
    \centering
    \includegraphics[width=0.9\columnwidth]{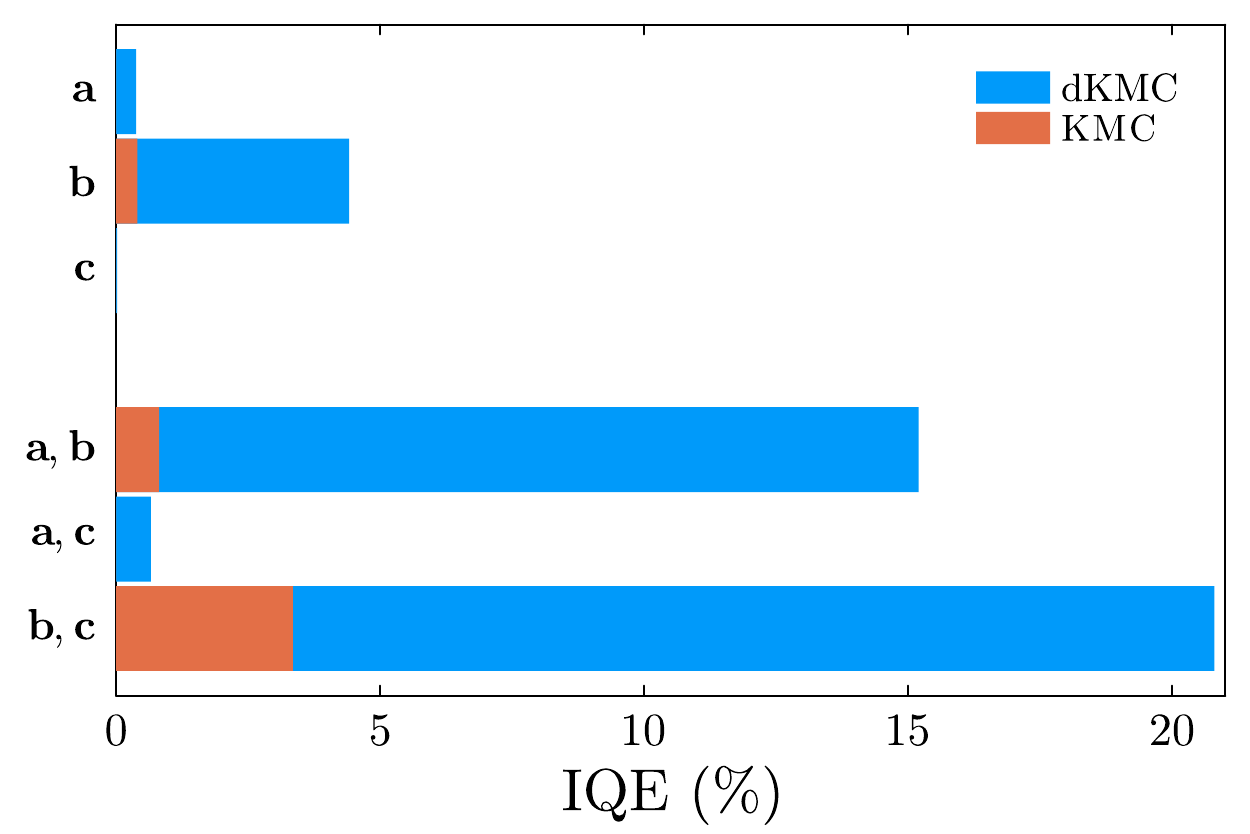}
    \caption{\textbf{Internal quantum efficiency (IQE)} of charge generation in neat Y6, calculated using classical KMC and dKMC. Included are results for lattices with different dimensionality where the unit cell is replicated in the direction of one or more of the primitive lattice vectors \vec{a}, \vec{b}, and \vec{c}. In all cases, when delocalisation is included using dKMC, the charge generation is significantly more efficient compared to that calculated using classical KMC.
    }
    \label{fig:IQE}
\end{figure}

\begin{table}
    \centering
	\begin{tabular}{llll}
		\toprule
		1D & $\mathrm{IQE}_\mathrm{KMC}$~(\%) & $\mathrm{IQE}_\mathrm{dKMC}$~(\%) & $ \dfrac{\mathrm{IQE}_\mathrm{dKMC}}{\mathrm{IQE}_\mathrm{KMC}}$ \\
		\midrule
		\vec{a} & \num{0.0020\pm0.0014} & \num{0.38\pm0.07}  & \num{190\pm140}\\
		\vec{b} & \num{0.40\pm0.04} & \num{4.41\pm0.18} & \num{10.9\pm1.3}\\
		\vec{c} &  \num{0.0010\pm0.0010} & \num{0.022\pm0.010} & \num{22\pm24} \\
        \midrule
		2D & & & \\
	    \midrule
		\vec{a},\vec{b} & \num{0.81\pm0.06} & \num{15.2\pm1.1} & \num{18\pm2}\\
		\vec{a},\vec{c} & \num{0.0030\pm0.0017} & \num{0.66\pm0.09} & \num{220\pm130} \\
        \vec{b},\vec{c} & \num{3.35\pm 0.13} & \num{20.8\pm0.4} & \num{6.2\pm0.3} \\
        \midrule
        3D & & & \\
	    \midrule
		\vec{a},\vec{b},\vec{c} & \num{5.49\pm0.14} &  & \\
        \bottomrule
	\end{tabular}
	\caption{\textbf{Internal quantum efficiency (IQE)} of charge generation in neat Y6, calculated using classical KMC and dKMC. Each row represents a lattice with different dimensionality, where the unit cell is replicated in the direction of one or more of the primitive lattice vectors \vec{a}, \vec{b}, or \vec{c}. Delocalisation significantly improves the charge generation in neat Y6, as quantified by the delocalisation enhancement in the right-most column. The uncertainties are the standard errors of the mean.} 
	\label{tab:IQE}
\end{table}

\begin{figure*}
    \centering
    \includegraphics[width=\textwidth]{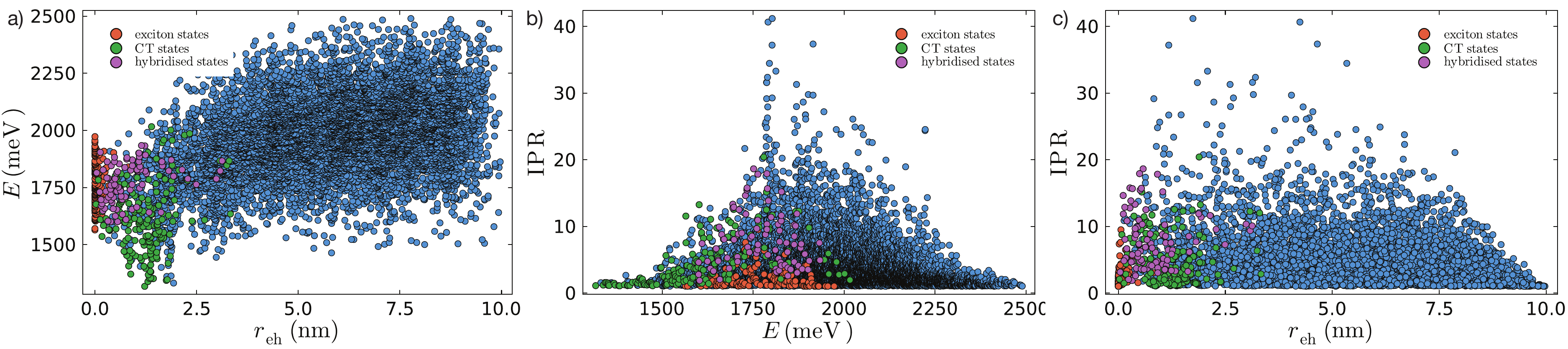}
    \caption{\textbf{Characteristics of delocalised polaron states.} Relationship between the energy $E$, inverse participation ratio~$\mathrm{IPR}$, and electron-hole separation $r_\mathrm{eh}$ of polaron states of Y6. The states are found by diagonalising a representative 2D $\tilde{H}_\mathrm{S}$ of Y6 with the unit cell replicated in the \vec{b} and \vec{c} directions. Some classes of polaron states are highlighted: exciton states are those with >75\% of their population on exciton site-pairs; CT states are those that have a CT recombination rate of >10\% of the maximum CT recombination rate; and hybridised states are those that have between 25\% and 75\% of their population on exciton site-pairs, with the remainder on site-pairs with greater electron-hole separation. 
    }
    \label{fig:polaron_states}
\end{figure*}

\Cref{tab:IQE} summarises our two-particle charge-generation simulations of the IQE in neat Y6, predicted using KMC and dKMC on lattices with different dimensionality. 

Our results show that delocalisation significantly improves the charge generation efficiency in neat Y6, with the delocalisation enhancement $\mathrm{IQE}_\mathrm{dKMC}/\mathrm{IQE}_\mathrm{KMC}=\num{6.0\pm0.5}$ in 2D. Like mobilities, IQEs are significantly higher on higher-dimensional lattices due to increased delocalisation and entropic effects. Although 3D calculations are currently intractable, we expect that the IQE would be even higher in 3D.

A direct comparison of IQEs with experiment is difficult, with a wide range reported in neat Y6~\cite{Price2022,Zhou2022,Mcanally2025}. 
IQEs can be measured both spectroscopically on films and through device measurements, with the two approaches having complementary strengths.
Spectroscopic measurements, using transient absorption spectroscopy, identified peaks implicated in charge generation~\cite{Price2022,Zhou2022,Giannini2024}, with the corresponding yields of 60--90\%~\cite{Price2022}, 29.3\%~\cite{Zhou2022}, and 60\%~\cite{Giannini2024}. It was argued that the peaks may not correspond to free charge carriers because of their short lifetime~\cite{Giannini2024}. However, a short lifetime is to be expected even for free charges in a neat material because of their high mobility and the absence of interfacial energetic offsets that would hinder recombination. The attribution of the peaks to free charges~\cite{Price2022,Zhou2022} is supported by the intensity dependence of the photoluminescence quantum yield~\cite{Price2022}, which indicates bimolecular recombination. Device measurements corroborate this conclusion, although care is needed in interpreting them because device IQEs can be affected by other processes, such as transport losses or undesired exciton dissociation at an interface between the active layer and a charge transport layer. Recent device work ruled out charge generation at transport-layer interfaces and ensured that IQE losses were not caused by poor extraction, finding the IQE to be 25\%~\cite{Mcanally2025}. Based on these values, we estimate the experimental range of neat Y6 IQE to be 25--60\%.

We find that including delocalisation can explain the efficient charge generation observed in neat Y6. Classical KMC significantly underpredicts the IQE, with a maximum of only \num{5.49\pm0.14}\% in 3D. By including delocalisation, dKMC increases the IQE to \num{20.1\pm0.5}\%, even though our calculations are limited to 2D. Therefore, delocalisation significantly improves the agreement with the experimental IQE range of 25--60\%. We expect this agreement will improve further once 3D dKMC simulations become computationally feasible.

Our results on Y6 are consistent with our previous mechanistic results on the role of delocalisation in charge generation~\cite{Balzer2024}. We previously established that delocalisation improves charge generation in a twofold way: by enabling charges and excitons to hop further and faster and by enabling the formation of hybridised states, those with both excitonic and separated character, which mediate the transfer of charges from excitons to separated charges~\cite{Balzer2024}. In \cref{fig:polaron_states}, we show that these hybridised states form in Y6, along with their properties.

\section{Discussion}
Our results demonstrate that dKMC offers a unified theoretical explanation of the remarkable transport and charge-generation properties of neat Y6. This work is the first parameterisation of dKMC based on atomistic calculations and experimental inputs, with no free parameters, and the calculated exciton diffusion coefficients, carrier mobilities, and IQEs are all consistent with experimental values. By contrast, classical KMC systematically underestimates each of these quantities, sometimes by more than an order of magnitude.

The large delocalisation enhancements determined by dKMC are the result of only modest delocalisation. \Cref{tab:transport} shows that the mean IPR of both carriers and excitons is only 2--3 molecules. In addition, these mean values overestimate the typical IPRs of the states that are thermally accessible or visited during the dynamics. 

A further key outcome is that high-dimensional transport is essential for describing Y6. Even though Y6 is anisotropic, modelling it as a one-dimensional stack along the $\pi$-$\pi$ stacking axis significantly underestimates delocalisation, mobilities, and IQEs. Extending the lattice to two dimensions increases the IPRs, carrier mobilities and the IQE. For hole mobilities, this is also true when extending the lattice to three dimensions, and we expect this additional improvement to hold for other carriers and the IQE as well. 

These conclusions are possible because dKMC strikes a favourable balance between cost and complexity. Although it includes a quantum-mechanical description of partially delocalised polarons, including two-particle electron–hole dynamics, it uses computational approaches that allow it to scale to mesoscopic lattices of millions of molecules and long simulation times, making it well suited for simulating disordered OPVs. Therefore, it complements approaches such as multiconfiguration time-dependent Hartree (MCTDH)~\cite{Tamura2013,Huix-Rotllant2015,Polkehn2018}, excitonic state-based surface-hopping (X-SH)~\cite{Peng2022,Ivanovic2025} and mesoHOPS~\cite{Citty2024,Varvelo2021}, which provide more accurate and fully quantum-mechanical descriptions of coupled electronic-vibrational dynamics, but at a cost that restricts them to smaller systems.

This work also establishes dKMC as a tool for computational screening. Our workflow for computing the necessary parameters can be used to systematically vary molecular properties to determine their effect on transport, IQEs, and other measurable properties.

dKMC can be further developed and optimised in several directions. Methodological improvements are likely to further reduce the computational cost and extend all of the calculations in this work to 3D. Additionally, some of the approximations could be relaxed, albeit at additional computational cost; for example, this could involve incorporating the full variational polaron transformation~\cite{Pollock2013,Jang2022} for a more accurate treatment of the slow bath modes. Finally, parameterised dKMC could be incorporated into multiscale, device-level simulations, where dKMC (or its simplification jKMC~\cite{Willson2023,Willson2023_2})  could be used to parameterise drift-diffusion simulations for calculations of device properties.

In conclusion, we have parameterised dKMC using atomistic calculations and experimental measurements to show that delocalisation can explain the high performance of neat Y6 OPVs. Delocalisation significantly improves the transport of electrons, holes, and excitons in Y6, finally reconciling theoretical predictions with high experimental mobilities and exciton diffusion coefficients. Delocalisation also substantially improves the IQE, explaining the surprisingly efficient devices made from neat Y6. These results demonstrate that efficient charge generation in neat Y6 is possible without the need for energy gradients, grain boundaries, defects, or high dielectric constants. The success of dKMC on the first material for which it was parameterised demonstrates its potential for realistic simulations of organic semiconductors. Because of its speed and accuracy, we anticipate that dKMC will further contribute to our understanding of the behaviour of increasingly higher-performance OPVs.

\begin{acknowledgments}
D.B. and I.K. were supported by a Westpac Scholars Trust Future Leaders Scholarship, the Australian Research Council (DP220103584), and the Australian Government Research Training Program. P.A.H. acknowledges support from the Marsden Fund and an MBIE Catalyst e-ASIA grant (RSCHTRUSTVIC2449). G.R.W. acknowledges the MacDiarmid Institute for Research Assistant funding and the JSPS Postdoctoral Fellowship Program for postdoctoral funding. We were supported by computational resources from the National Computational Infrastructure (Gadi), the University of Sydney Informatics Hub (Artemis), and the Victoria University of Wellington high-performance-computing cluster (Rāpoi).
\end{acknowledgments}

\bibliography{bib}

\end{document}


\title{Supporting Information: Delocalisation explains efficient transport and charge generation in neat Y6 organic photovoltaics}

\author{Daniel Balzer}
\affiliation{School of Chemistry and University of Sydney Nano Institute, University of Sydney, NSW 2006, Australia}

\author{Paul A. Hume}
\affiliation{School of Chemical and Physical Sciences, Victoria University of Wellington, Wellington, New Zealand}
\affiliation{MacDiarmid Institute for Advanced Materials and Nanotechnology, Wellington, New Zealand}
\affiliation{The Dodd-Walls Centre for Photonic and Quantum Technologies, Dunedin 9016, New Zealand}

\author{Geoffrey R. Weal}
\affiliation{School of Chemical and Physical Sciences, Victoria University of Wellington, Wellington, New Zealand}
\affiliation{MacDiarmid Institute for Advanced Materials and Nanotechnology, Wellington, New Zealand}
\affiliation{Institute for Integrated Cell-Material Sciences (iCeMS), Kyoto University, Kyoto, Japan}

\author{Justin M. Hodgkiss}
\affiliation{School of Chemical and Physical Sciences, Victoria University of Wellington, Wellington, New Zealand}
\affiliation{MacDiarmid Institute for Advanced Materials and Nanotechnology, Wellington, New Zealand}

\author{Ivan Kassal}
\affiliation{School of Chemistry and University of Sydney Nano Institute, University of Sydney, NSW 2006, Australia}

\maketitle

\setcounter{section}{0}
\renewcommand{\thesection}{S\arabic{section}}%
\setcounter{equation}{0}
\renewcommand{\theequation}{S\arabic{equation}}%
\setcounter{figure}{0}
\renewcommand{\thefigure}{S\arabic{figure}}%
\setcounter{table}{0}
\renewcommand{\thetable}{S\arabic{table}}%
\setcounter{algorithm}{0}
\renewcommand{\thealgorithm}{S\arabic{algorithm}}%

\section{Parameters}

\begin{table*}[h]
	\centering
    \renewcommand{\arraystretch}{1.2}
    \normalsize
	\begin{tabular}{lll}
		\toprule
        Parameter & Description & Values \\
        \midrule
        $d$ & Dimension & 1--3\\
		$N$ & Number of unit cells along each dimension & 100\\
        $[a, b, c]$ & Lattice constants & [1.37272, 1.96561, 2.97056]~nm \\
        $[\alpha,\beta,\gamma]$ & Lattice angles & [\ang{102.450}, \ang{92.677}, \ang{96.570}] \\
        & Fractional coordinates of molecule 1 & [0.380, 0.684, 0.713] \\ 
        & Fractional coordinates of molecule 2 & [0.123, 0.070, 0.707] \\
        & Fractional coordinates of molecule 3 & [0.619, 0.328, 0.300] \\
        & Fractional coordinates of molecule 4 & [0.877, 0.930, 0.293]\\
        $\sigma_\mathrm{e}$ & Electron disorder & \qty{103}{meV}\\
        $\sigma_\mathrm{h}$ & Hole disorder & \qty{100}{meV} \\
        $\sigma_\mathrm{xc}$ & Exciton disorder & \qty{50}{meV}\\
        $T$ & Temperature & \qty{300}{K} \\
        $L_\mathrm{xc}$ & Exciton lifetime & $\num{2.6e-10}~\mathrm{s}^{-1}$ \\
        $R_\mathrm{vib}$ & Vibrational dephasing rate & $\num{e12}~\mathrm{s}^{-1}$ \\
        $\Delta$ & Representative dimer splitting & \qty{0}{meV} \\ 
        $a_\mathrm{dKMC}$ & dKMC accuracy & 0.99 \\
        $r_\mathrm{sep}$ & Separation cutoff & \qty{5}{nm}\\ 
        $n_\mathrm{hops}$ & Maximum number of hops & 10000 \\
        $n_\mathrm{iter}$ & Number of simulation landscapes & 1000\\
        $n_\mathrm{traj}$ & Number of trajectories on each landscape &  100 in 1D, 10 in 2D, 10 in 3D\\
		\bottomrule
	\end{tabular}
	\caption{
	\textbf{Parameter values.} Default values used for parameters in dKMC simulations of Y6, unless stated otherwise.}
	\label{tab:parameters}
\end{table*}

\newpage
\section{Parameterising the Coulomb interaction}

\begin{figure*}[h]
    \centering
    \includegraphics[width=0.6\textwidth]{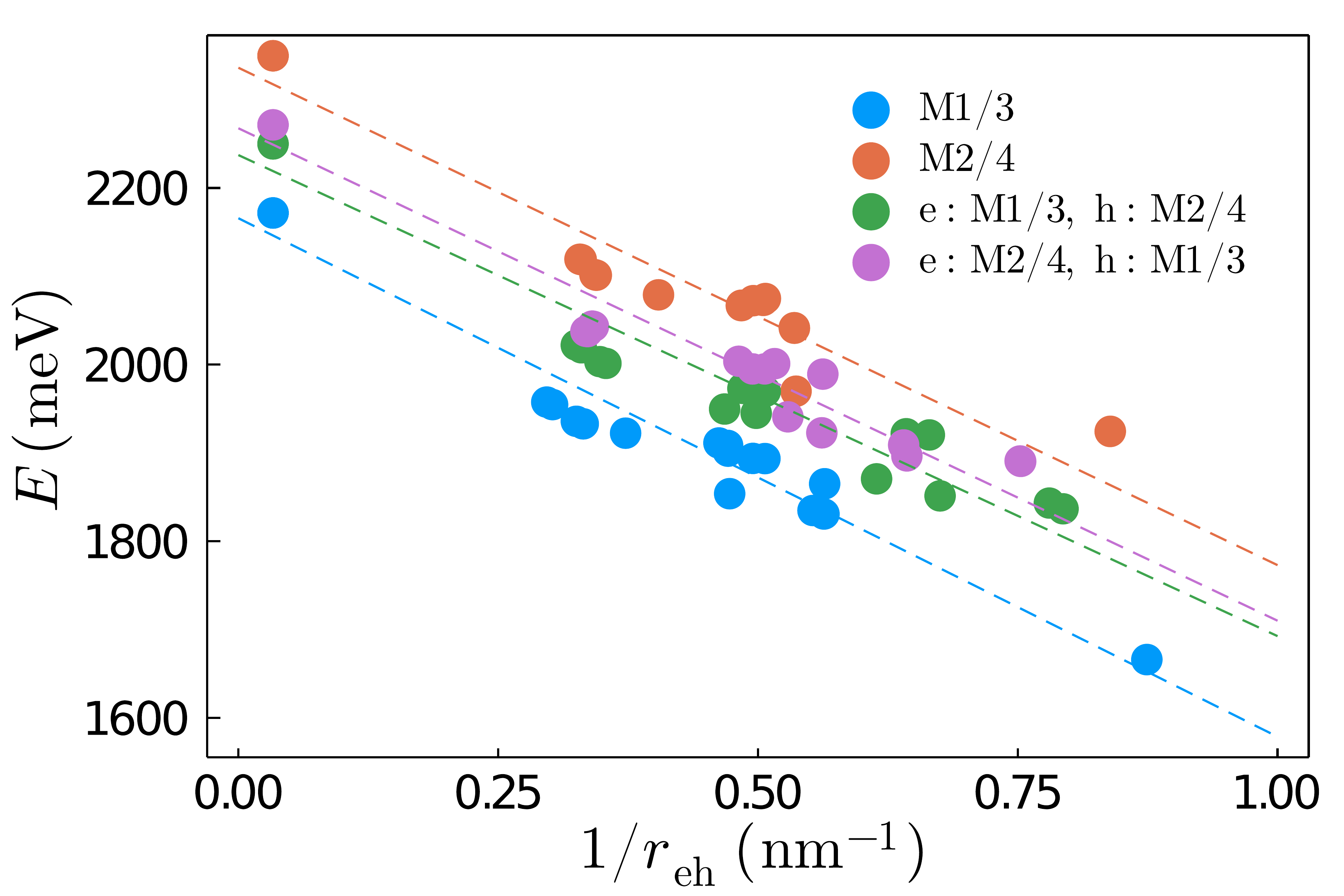}
    \caption{\textbf{Parameterising the Coulomb interaction.} The energy of molecular pairs as a function of the inverse electron-hole separation. As the gradients for each category of molecular pair are similar, we parameterise the long-range Coulombic interaction using the mean of the gradients of the linear fits, giving $U(r_\mathrm{eh})=-\qty{563}{meV.nm}/r_\mathrm{eh}$, equivalent to a relative permittivity of $\epsilon_r=2.56$.
    }
    \label{fig:coulomb_interaction}
\end{figure*}

\newpage
\section{Parameterising spectral densities}

\begin{figure*}[h]
    \centering
    \includegraphics[width=\textwidth]{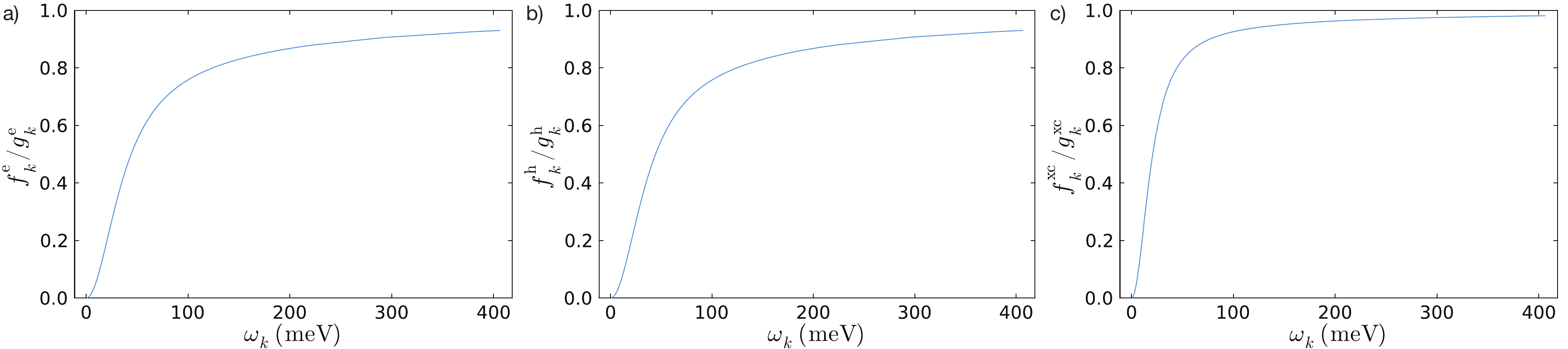}
    \caption{\textbf{Variational optimisation.} The full variational curve of $f_k
    ^z(\omega)/g_k^z(\omega)$ for \textbf{(a)}~electrons, \textbf{(b)}~holes, and \textbf{(c)}~excitons, produced from the variational optimisation of the system-bath couplings of every bath mode on a representative dimer.
    }
    \label{fig:variational_curves}
\end{figure*}

\begin{figure*}[h]
    \centering
    \includegraphics[width=\textwidth]{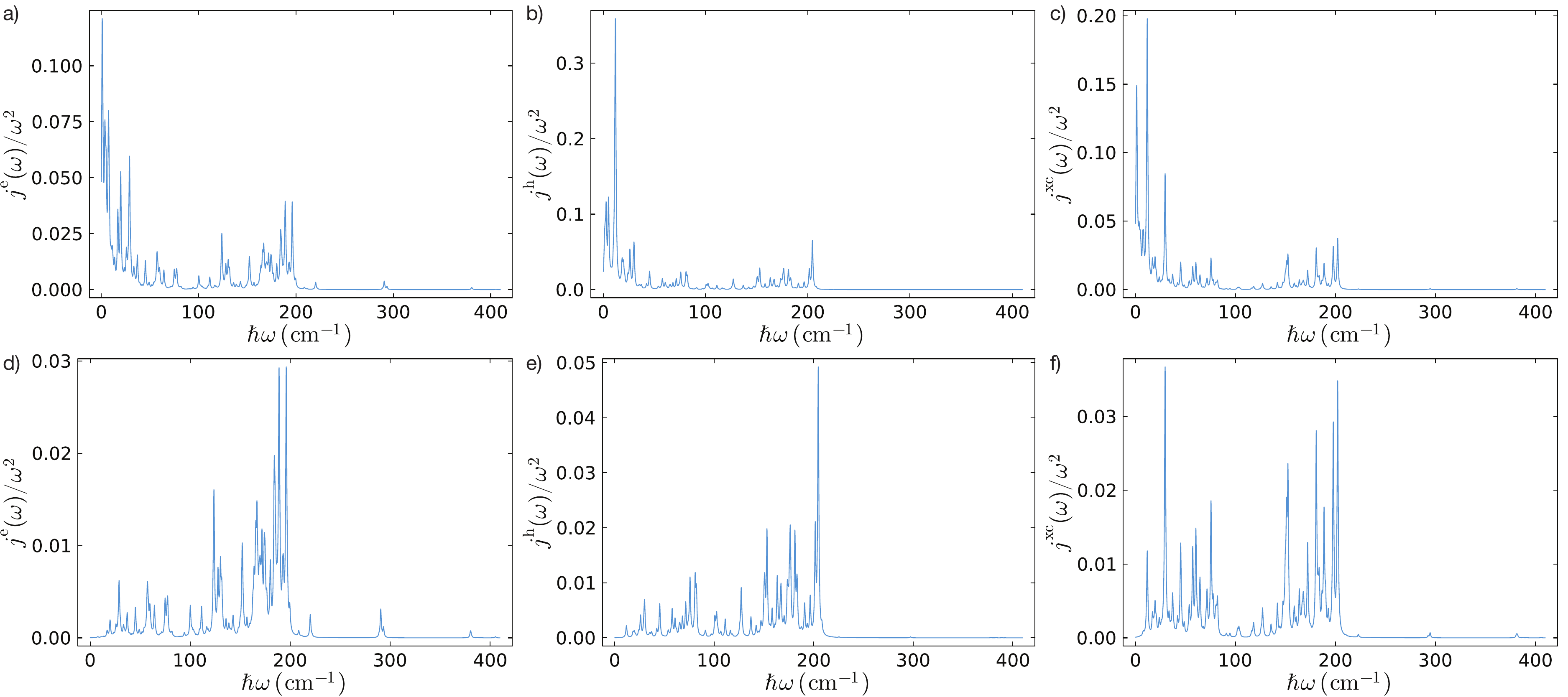}
    \caption{\textbf{Spectral densities.} The continuous spectral densities $j^z(\omega)$ (plotted as $j^z(\omega)/\omega^2$) for \textbf{(a)} electrons, \textbf{(b)} holes, and \textbf{(c)} excitons produced from the Huang-Rhys factors of the normal modes, and \textbf{(d--f)} the corresponding spectral densities following the partial mode displacement after the variational optimisation of system-bath couplings on a representative dimer. The partially displaced spectral densities are found by multiplying the fully displaced ones by $f^z_k(\omega)^2/g^z_k(\omega)^2$, the square of the corresponding variational curve in \cref{fig:variational_curves}. As a result, in the partially displaced cases, the slow, low-frequency modes contribute less to the spectral density.
    }
    \label{fig:spectral_densities}
\end{figure*}
